\documentclass[sigconf,authorversion]{acmart}

\usepackage{graphicx}
\usepackage{subcaption}
\usepackage{adjustbox}
\usepackage[all]{nowidow}

\AtBeginDocument{%
  \providecommand\BibTeX{{%
    \normalfont B\kern-0.5em{\scshape i\kern-0.25em b}\kern-0.8em\TeX}}}

\copyrightyear{2022}
\acmYear{2022}
\setcopyright{acmcopyright}\acmConference[GAS'22]{6th International ICSE Workshop on Games and Software Engineering: Engineering fun, inspiration, and motivation}{May 20, 2022}{Pittsburgh, PA, USA}
\acmBooktitle{6th International ICSE Workshop on Games and Software Engineering: Engineering fun, inspiration, and motivation (GAS'22), May 20, 2022, Pittsburgh, PA, USA}
\acmPrice{15.00}
\acmDOI{10.1145/3524494.3527625}
\acmISBN{978-1-4503-9293-8/22/05}
\begin{document}

\title{Towards Self-Adaptive Game Logic}

\author{Erik M. Fredericks}
%\email{webmaster@marysville-ohio.com}
\affiliation{%
  \institution{Grand Valley State University}
  \streetaddress{1 Campus Dr.}
  \city{Allendale}
  \state{Michigan}
  \country{USA}
  \postcode{49401}
}
\email{frederer@gvsu.edu}

\author{Byron DeVries}
\affiliation{%
  \institution{Grand Valley State University}
  \streetaddress{1 Campus Dr.}
  \city{Allendale}
  \state{Michigan}
  \country{USA}
  \postcode{49401}
}
\email{devrieby@gvsu.edu}

\author{Jared M. Moore}
\affiliation{%
  \institution{Grand Valley State University}
  \streetaddress{1 Campus Dr.}
  \city{Allendale}
  \state{Michigan}
  \country{USA}
  \postcode{49401}
}
\email{moorejar@gvsu.edu}

\renewcommand{\shortauthors}{Fredericks et al.}

\begin{abstract}
 Self-adaptive systems (SAS) can reconfigure at run time in response to changing situations to express acceptable behaviors in the face of uncertainty.  With respect to game design, such situations may include user input, emergent behaviors, performance concerns, and combinations thereof.  Typically an SAS is modeled as a feedback loop that functions within an existing system, with operations including monitoring, analyzing, planning, and executing (i.e., MAPE-K) to enable online reconfiguration.  This paper presents a conceptual approach for extending software engineering artifacts to be self-adaptive within the context of game design.  We have modified a game developed for creative coding education to include a MAPE-K self-adaptive feedback loop, comprising run-time adaptation capabilities and the software artifacts required to support adaptation.  %Initial results demonstrate   
\end{abstract}





%   \includegraphics[width=\textwidth]{feesh.PNG}
%   \caption{\texttt{feesh} screenshot (update for anonymity).}
%   \Description{wat is this.}

\maketitle

\section{Introduction}

Games are generally structured towards being reactive to the player to provide an immersive experience. For example, difficulty may scale as the player's character (PC) grows in strength or the environment and/or story may change as the PC makes in-game decisions.  Such updates may be considered to be adaptations to enrich gameplay.  In the case of an arcade-style game (i.e., where high score is the goal), such changes include reactions to the player's skill level, managing performance concerns, and optimizing overall game flow.  This paper introduces a technique for incorporating software artifacts (e.g., requirements, goals, etc.) as first-class citizens in an adaptive gameplay loop.  We apply this technique to our existing open-source game project to demonstrate its effectiveness.

Self-adaptive systems (SAS) are an approach for enabling reconfiguration at run time in systems facing uncertainty~\cite{kephart.2003,mckinley.2004}, where the MAPE-K feedback loop~\cite{kephart.2003} is one common technique for enabling adaptation.  For instance, a system may change its configuration, algorithm, or decision-making engine as adversity manifests from combinations of system or environmental uncertainty (e.g., human interaction, unexpected external conditions, etc.).  
In terms of games, such configurations may include the number and intelligence of enemy characters, distribution of item pickups in relation to player level, and maintaining an appropriate frame rate as the number of instantiated objects increase.
Moreover, software engineering techniques such as goal modeling~\cite{lamsweerde.1998} and run-time requirements monitoring~\cite{degrandis.2009} can enable introspection that supports self-reconfiguration strategies~\cite{bowers.2020}, however these self-reconfiguration techniques have not been intrinsically applied to a game loop.

This paper describes our application of a MAPE-K feedback loop to a game designed as a creative coding learning experience for students.  We generated the necessary software artifacts (i.e., a goal model~\cite{lamsweerde.2009} with utility functions~\cite{degrandis.2009}), updated existing source code with a self-adaptive overlay, and added instrumentation for monitoring the performance of the overall system with respect to our defined metrics.    

% expr?

Our initial results demonstrate the feasibility of applying self-adaptation to a gameplay loop. 
The remainder of this paper is structured as follows.  Section~\ref{sec:bg} discusses relevant background information on creative coding (focusing on our motivating example), SASs, and goal modeling.
Section~\ref{sec:appr} then discusses our approach for modeling a game and its software artifacts as a self-adaptive feedback loop and presents our initial results.  Lastly, Section~\ref{sec:discussion} summarizes our results and presents future directions.

\section{Background and Related Work}%Motivating ExampleBackground}
\label{sec:bg}

This section describes our motivating example, SASs, and goal modeling.  For each sub-section we also highlight related work.

%This section introduces relevant background information on creative coding, our motivating example, SASs, and goal modeling.

\subsection{Creative Coding (Motivating Example)}

Creative coding is a form of programming in which the goal is generally considered to be some form of generative artwork or video game, typically created via abstractions over existing programming languages~\cite{greenberg.2007}.  Processing (Java) and p5.js (JavaScript)\footnote{See \url{https://processing.org/} and \url{https://p5js.org/}, respectively.} are two such abstractions that facilitate rapid development of creative coding applications.  As such, creative coding can additionally be used to introduce students to high-level programming and game design concepts while minimizing the overhead and/or learning curve of graphics libraries (e.g., OpenGL, WebGL, SDL, etc.).  Creative coding, as applied to this project, was intended to serve as a demonstration for undergraduates on game design concepts.  We next describe \texttt{feesh}, our motivating example.

\noindent \textbf{Motivating example - \texttt{feesh}}:  \texttt{feesh} is a relatively simple web-based game where the intention is to grow by eating smaller objects.  This project was written in p5.js and can run in any browser without the need for a server.  Additionally, the source code is available on GitHub.\footnote{\url{https://github.com/efredericks/feesh}.}

\begin{figure}[htb!]
\includegraphics[width=3.3in]{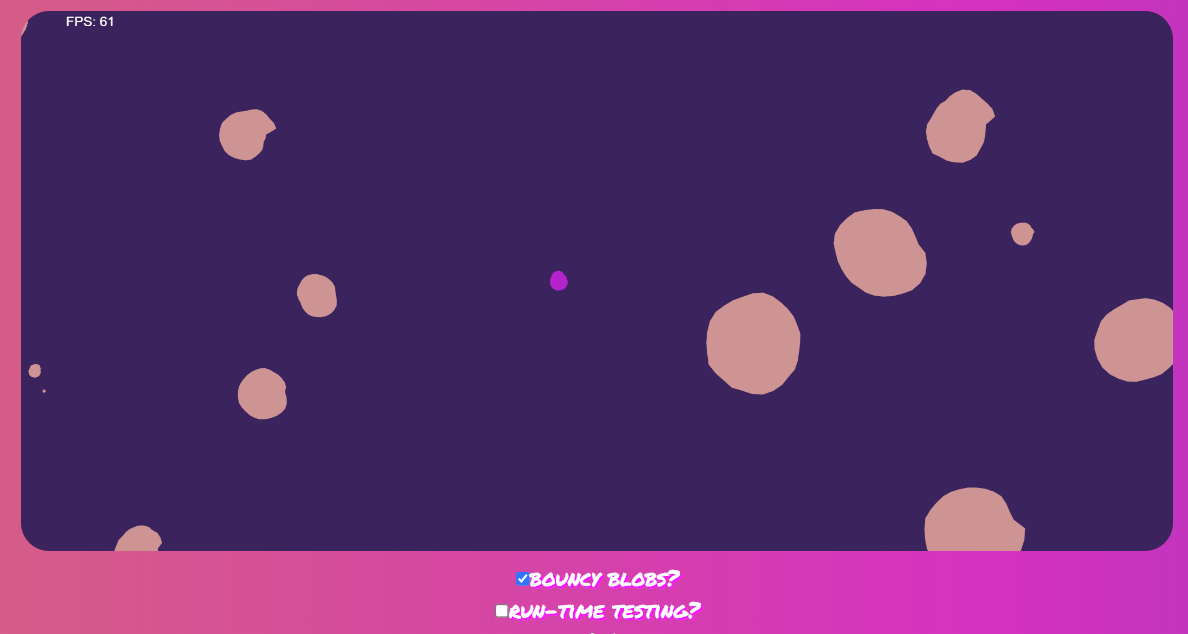}
\caption{Screenshot of feesh gameplay. The pink blob is the player and the tan blobs are enemies of varying size.  The checkboxes on the bottom can change game parameters at run time.}
\label{f:feesh}
\end{figure}

The gameplay loop for \text{feesh} is based on Fishy, a classic Flash game where the intent is to avoid large fish and eat smaller fish.\footnote{See \url{https://www.silvergames.com/en/fishy}.}  The intent of \texttt{feesh} was to replicate the core mechanics of Fishy while introducing self-adaptation into the core gameplay loop.  For additional processing complexity, all objects are formed via noise loops to give a ``wobbly'' impression, based upon the \textit{TheCodingTrain}'s Blobby! video~\cite{shiffman.2016}.  Collision detection is simplified to be circle-circle collision between all entities.

Creative coding is typically structured around introducing students to computer science concepts via generative art or simplified game design as a way for students to graphically visualize difficult problems~\cite{peppler.2005, bergstrom.2015}, with popular textbooks including \textit{The Nature of Code}~\cite{shiffman.2012} and \textit{Generative Art: A Practical Guide using Processing}~\cite{pearson.2011}.  Recently, creative coding has been applied to Internet of Things applications to further express art in real-world settings~\cite{vestergaard.2017}.  However, many papers centered around creative coding focus on teaching coding concepts, whereas we posit learning can be extended to additionally comprise complex software engineering concepts (where this argument will be explored in future works).  Similarly, Bucchiarone \textit{et al}. discuss gamification as a means for enhancing user engagement~\cite{bucchiarone.2019,bucchiarone.2021}, however for the purposes of this paper we aim to enhance gameplay via self-adaptation in contrast to adding gameplay elements an existing application.

\subsection{Self-Adaptive Systems}

SASs provide an approach for self-reconfiguring (e.g., configuration, algorithm, etc.) at run time~\cite{mckinley.2004,oreizy.1999} in response to changing environmental and system conditions to continuously satisfy key objectives~\cite{cheng.2009,whittle.2009}.  The requirements for the system itself may also change over time, potentially necessitating updates in terms of patches, bug fixes, or configuration change.  For the purposes of this paper we consider an SAS to comprise a set of configurable states connected via adaptive logic~\cite{zhang.2006}.  While there exist multiple approaches for enabling self-adaptation, we will follow the MAPE-K (Monitor-Analyze-Plan-Execute-Knowledge) feedback architecture~\cite{kephart.2003}.  We next describe our motivating example in the context of MAPE-K to highlight each aspect of its architecture.

\noindent \textbf{Monitor}:  An SAS must monitor itself and its environment for decision-making purposes.  We specified key metrics as relevant to the adaptive properties of our game engine, with \textit{internal software monitors} being responsible for watching the state of each metric.  Specifically, we manually identified \textit{which} aspects of our game engine were amenable to adaptation (e.g., player size, enemy count) and which metrics \textit{might} impact those aspects (e.g., FPS, execution time)  A subset of these metrics, along with thresholds and reconfiguration strategies, are listed as follows in Table~\ref{tab:mon}.  Note, all values found in Table~\ref{tab:mon} were empirically derived.

\begin{table*}
  \caption{Monitored Metrics and Thresholds (N/A reconfigurations are only monitors).}
  \label{tab:mon}
  \begin{tabular}{lcll}
    \toprule
    Monitor & Threshold & Reconfiguration Strategy & Affected Goal(s)\\
    \midrule
    Frame rate (FPS) & >=30 & [Reduce enemy count, Disable enemy-enemy collision] & (B), (D), (E)\\
    Playability & Maximize & [Reduce player size, Reduce enemy count] & (E), (F)\\
    Score & Maximize & N/A & (H) \\
    Number of enemies on screen & Maximize & N/A & (B), (E) \\%[Reduce enemies]\\
    Enemy-enemy collision enabled? & T|F & N/A & (B), (D)\\
    Execution time & Maximize & [Reduce player size, Reduce enemy count] & (A), (B), (C), (D), (E), (F)\\
    Random event & $2\%$ chance & [Increase enemy count] & (A), (C), (E)\\
  \bottomrule
\end{tabular}
\end{table*}

Note that, for the purposes of this paper we are mainly focusing on features that directly impact the player during gameplay, however for future work we could examine past-game history (e.g., player behaviors, enemy encounters, etc.) to determine if adaptations are warranted.

\noindent \textbf{Analyze}: Each monitored attribute is then analyzed to determine if a violation has occurred or a threshold has been exceeded, thus necessitating a reconfiguration.\footnote{In other systems, continuous learning may be implemented to make predictions, however self-reconfiguration in \texttt{feesh} is reactive.}  For this project, we defined thresholds for each monitored attribute to determine if a reconfiguration is warranted.  For example, if the frame rate (i.e., frames per second, or \texttt{FPS}) falls below $30$ then an adaptation is necessary to ensure the player has a smooth experience.

\noindent \textbf{Plan}: Based on identified adaptation requirements the \textit{Plan} phase will select a reconfiguration strategy.  Depending on the complexity of the system and feedback loop, this phase may involve the generation of many possible adaptations.  
%For this project, we have tightly-coupled the \textit{Analyze} and \textit{Plan} phases to enable simple adaptation.  Specifically, if an adaptation is required then a correlated strategy will be selected.  
For example, when the \texttt{FPS} drops below $30$ the system will have the choice to remove the enemy-enemy collision (i.e., calculating collisions between enemies to induce ``bouncing'') or to remove a random number of enemies to reduce overall processing load.

\noindent \textbf{Execute}: The system will then execute the adaptation strategy based on monitored data and the decisions made in the \textit{Analysis} and \textit{Plan} phases.  In the case of \texttt{feesh}, self-adaptation can comprise changes in complexity (e.g., increase the number of enemies given acceptable monitored performance), reduce the size of entities (e.g., scale down player size to improve playability),
% update enemy intelligence (e.g., to make the game harder or simpler), 
or change collision detection between enemies (e.g., removing enemy-enemy collision to improve performance).  
The {Execute} phase then directly applies the reconfiguration as needed.

\noindent \textbf{Knowledge}: For simplicity, \textit{knowledge} of all aspects of the game and its self-adaptation mechanics are globally-accessible to all components of the MAPE-K loop.

\noindent \textbf{Uncertainty}: For the purposes of this paper, we will consider uncertainty to reflect human interaction (i.e., the player), the capabilities of the player's computer (e.g., browser choice, hardware specifications, frame rate, etc.), and random behaviors implemented within the \texttt{feesh} game engine.  The combination of these types of uncertainty can result in different gameplay experiences, and as such, \texttt{feesh} has been modeled as self-adaptive.  

Previously, the MAPE-K loop has been applied to a gaming application by Yamagata \textit{et al}.~\cite{yamagata.2019}, however their application focused mainly on the latency issue in multiplayer games, whereas our application focuses on including adaptation as part of the intrinsic game mechanics.  C{\'a}mara \textit{et al}. investigated how an SAS can optimize network latency via gamification and by modeling it as a stochastic multiplayer game~\cite{camara.2016}.   Other methods of adaptation, among many, include the use of dynamic software product lines to update behaviors at run time~\cite{ayala.2021}, Bayesian optimization with fuzzy systems~\cite{pirovano.2012}, and artificial intelligence~\cite{swiechowski.2013}.

\subsection{Goal Modeling}

\begin{figure}[htb!]
\includegraphics[width=3.3in]{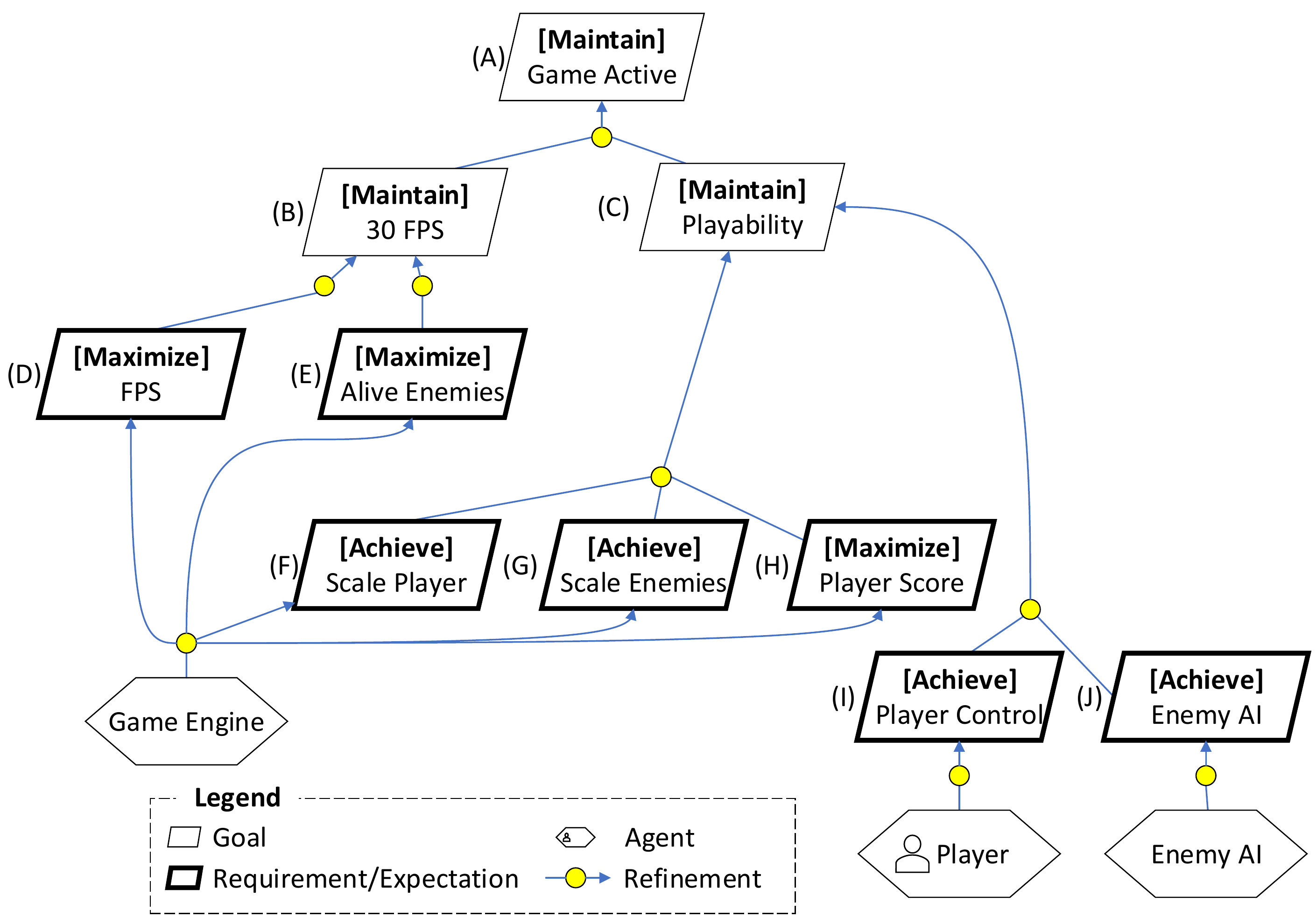}%goal-model.pdf}
\caption{Initial goal model of \texttt{feesh} application.}
\label{fig:goal-model}
\end{figure}

Goal-oriented requirements engineering (GORE) uses goals (i.e., high-level objectives and/or requirements) to model system behaviors and objectives~\cite{lamsweerde.2009}.  GORE models a system-to-be via an acyclic graph comprising goals, refinements, requirements/expectations (i.e., leaf-level goals), and agents.  Knowledge Acquisition in Automated Specification (KAOS) extends GORE via additional AND- and OR-refinements~\cite{lamsweerde.2009, dardenne.1993} (e.g., Goals (A) and (B) in Figure~\ref{fig:goal-model}, respectively)..  AND-refined goals are satisfied only when \textit{all} subgoals are satisfied and OR-refined goals are satisfied when at minimum \textit{one} of its subgoals are satisfied.  Figure~\ref{fig:goal-model} presents a KAOS~\cite{lamsweerde.2009} goal model for \texttt{feesh}, where KAOS is one approach to goal modeling.\footnote{A similar approach for goal modeling is iStar~\cite{yu.1997}.}

The main goal of \texttt{feesh} is to extend time played (similar to an arcade game) (Goal (A)).  To satisfy this goal, the game must be at an acceptable frame rate (Goal (B)) while ensuring it remains playable (Goal (C)).  Goals (D) and (E) attempt to balance FPS by managing collision detection and the number of enemies on screen, respectively.  Goals (F), (G), and (H) support Goal (C) by keeping the player at a manageable size while ensuring the score continually increases (to maximize player satisfaction).  Moreover, Goals (I) and (J) additionally support (C) by enabling player and enemy control, respectively.  

Goal (A) represents an AND-satisfied goal (i.e., Goals (B) and (C) must be satisfied for (A) to be satisfied) and Goal (B) represents an OR-satisfied goal (i.e., either Goals (D) or (E) must be satisfied for (B) to be satisfied).  Moreover, goals designated as \texttt{Maintain} are considered to be \textit{invariant} and those designated as \texttt{Achieve} are considered as \textit{non-invariant}, where \texttt{invariant} goals cannot tolerate a violation and \texttt{non-invariant} goals can temporarily accept a transient violation.

Note that maximizing the number of enemies on screen while setting enemy-enemy collision to be enabled are competing concerns, as the collision detection algorithm implemented may significantly reduce frame rate. We would like to note that more ``intelligent'' collision detection algorithms can be implemented to minimize this issue (e.g., executed in parallel on a GPU, using local search/partitioning, etc.), however for the purposes of this paper a simpler implementation was preferred to demonstrate the issue.

We next describe utility functions, the approach we use for determining goal satisfaction at run time.

\noindent \textbf{Utility Functions}:  Utility functions are mathematical formulae that have been used to quantify software requirement satisfaction~\cite{degrandis.2009,ramirez.2011,walsh.2004}.  In the context of an SAS, a utility function can serve as a metric for deciding if a reconfiguration is necessary (e.g., in the event that a goal is violated or under-performing).  For example, a utility function can be derived for Goal (D) in Figure~\ref{fig:goal-model} as follows in Equation~\ref{eqn:utilB}:

\begin{equation}
   {util}_B = \begin{cases}
   1.0 \quad &\text{if} \, FPS \geq 40 \\
   f(x) \quad &\text{else if} \, FPS \geq 30 \\
   0.0 \quad &\text{else} \\
   \end{cases}
    \label{eqn:utilB}
\end{equation}

\noindent , where a value of $0.0$ indicates a violation, $1.0$ indicates satisfaction, and any value in between denotes the degree of \textit{satisficement}. For this project, each utility function yields a value normalized on $[0.0, 1.0]$ and additionally implements a threshold, where any value below that threshold is considered inadequate and requires adaptation to resolve.  In some cases (e.g., Goal (A), (G), (I), and (J)), the utility function is trivially true in that it always is $1.0$.  Values such as these are generally not useful as points of adaptation given that there is no change, whereas Goal (D) will change drastically over the course of the game execution.

\section{Approach}
\label{sec:appr}

This section describes how self-adaptation can be applied to a gameplay loop via our motivating example and further presents initial experimental results gathered during gameplay by the authors.

\subsection{Instrumentation}

We first developed a goal model of \texttt{feesh} as seen in Figure~\ref{fig:goal-model}.  For each goal we developed a utility function (e.g., Equations~\ref{eqn:utilB} and \ref{eqn:utilF} for examples) to measure the satisfaction of each goal at run time.  Additionally, we introduced adaptation mechanisms into \texttt{feesh} comprising software sensors (Monitoring) and logic for enabling self-reconfiguration (Analysis, Planning, Execution, Knowledge).  Table~\ref{tab:mon} provides the adaptations currently available in \texttt{feesh}, however we anticipate that additional adaptation mechanisms can be introduced in future work.  Each of these activities requires domain knowledge of the developer to (1) understand the underlying game logic/engine and know when/where software can be augmented with self-adaptive characteristics and (2) be able to derive new software artifacts and/or leverage existing software artifacts for use as first-class citizens at run time.

\subsection{Adaptive Gameplay}

\begin{figure}
\begin{adjustbox}{right}
   \begin{subfigure}{3.3in}%1.2\columnwidth}
      \includegraphics[width=\linewidth]{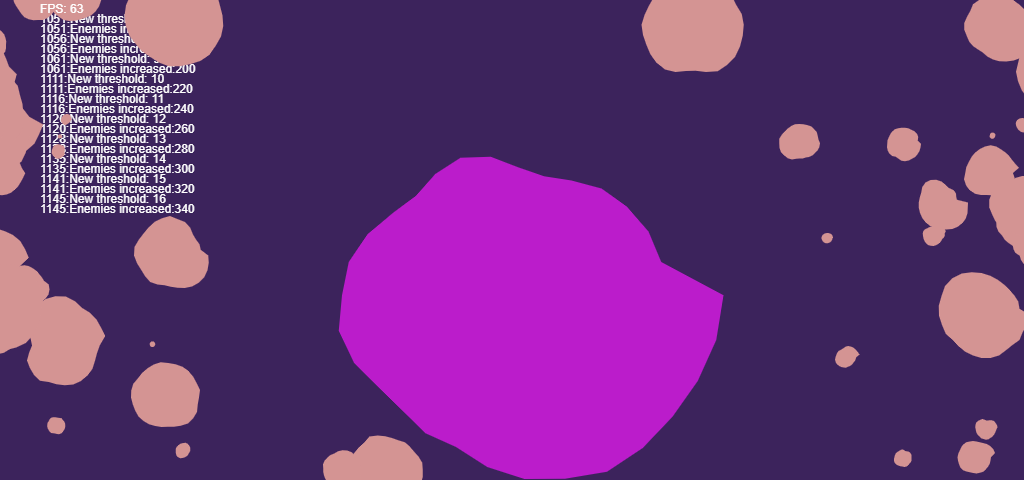}
      \caption{Sample of large player character that impedes on Goal (C).}
      \label{fig:reconfig1}
   \end{subfigure}
\end{adjustbox}

\vspace{0.2in}

\begin{adjustbox}{right}
   \begin{subfigure}{3.3in}%1.2\columnwidth}
      \includegraphics[width=\linewidth]{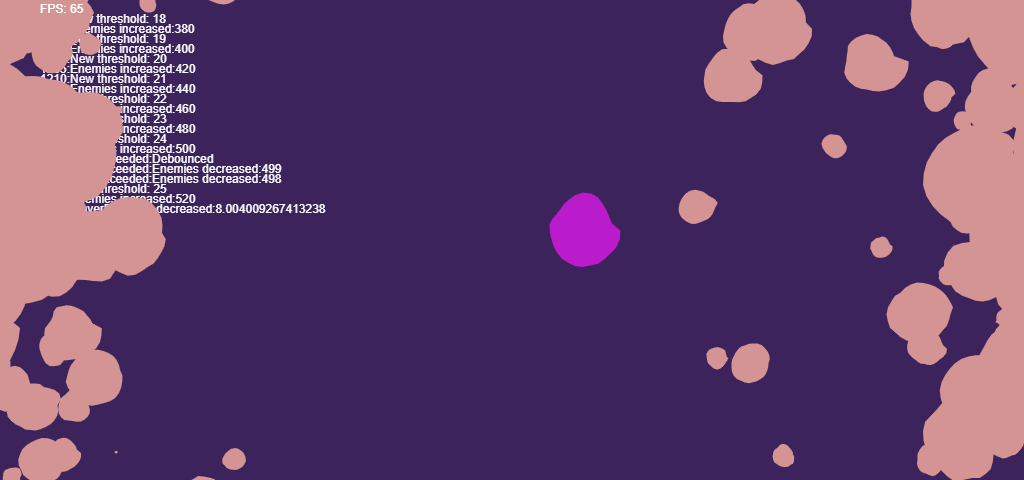}
      \caption{Player size reduced as result of MAPE-K adaptation.}
      \label{fig:reconfig2}
   \end{subfigure}
\end{adjustbox}
\caption{Reconfiguration resulting from Goal (C) violation.}\label{fig:reconfigs}
\end{figure}

We now present a sample use case for the MAPE-K loop within the context of the game.  For example, consider Goal (C) in Figure~\ref{fig:goal-model}.  The utility function for this goal can be inserted into the main gameplay loop as part of the \textit{Monitoring} phase.  In the event that this goal is considered violated (e.g., $playerSize$ $>$ $width/2$) then the adaptation mechanism will determine which reconfiguration is necessary for Goal (C) to be satisfied again (note: such goals are considered \textit{non-invariant} in that they can tolerate temporary failure; an \textit{invariant} goal cannot tolerate failure and as such the program would need to go to a safety state - e.g., shutdown or forcing user input).  Depending on the complexity of the Analysis and Planning phases, the SAS may generate one or many possible reconfiguration strategies and select the most appropriate course of action.  In the case of Goal (C), the system reduces $playerSize$ by $50\%$ (where this value was determined empirically to extend gameplay).  Moreover, if the $playerSize$ exceeds the width of the canvas (i.e., nothing else is visible other than the player) then the game is considered to be \textit{won} and the player is returned to the main menu.\footnote{With MAPE-K active, the \textit{won} state is not attainable unless if the player gains mass quicker than can be processed.}  Figures~\ref{fig:reconfig1} and \ref{fig:reconfig2} illustrate this adaptation in-game.
%Figure~\ref{} demonstrates the utility value of Goal (TBD) over the course of the game.

% \begin{figure}[htbp]
%   \centering
%   \includegraphics[width=3.2in]{feesh-large.png}
%   \caption{Sample of large player character that requires reduction.}
%   \label{fig:game1}
% \end{figure}
%         \hfill
% \begin{figure}[htbp]
%   \centering
%   \includegraphics[width=3.2in]{feesh-small.png}
%   \caption{Player size reduced as a result of MAPE-K adaptation.}
%   \label{fig:game2}
% \end{figure}

%\begin{figure}[H]

\subsection{Initial Experimental Results}

This section presents initial results on applying the MAPE-K feedback loop to \texttt{feesh}.   For the purposes of this experiment, all utility values are normalized between $[0.0, 1.0]$ and execution time relies on the \texttt{frameCount} variable provided by p5.js (i.e., integer ticks since program start).  \texttt{feesh} was run on a GitHub.io-hosted page (\url{https://efredericks.github.io/feesh/}) and accessed via Google Chrome (Version 97.0.4692.71 (Official Build) (64-bit)) on a modern laptop.  We performed two experimental treatments: MAPE-K enabled and MAPE-K disabled (i.e., \textit{Normal}).  For each treatment we performed $50$ replicates to establish statistical significance and we use the Wilcoxon-Mann-Whitney u-test with a significance (i.e., \textit{p}-value) threshold of $0.05$.  The results were gathered by the authors of this paper.  Additionally, we focus on Goal (F) to demonstrate feasibility of this approach, where we will explore additional metrics in future work.

\begin{figure}[h!]
\begin{adjustbox}{right}
   \begin{subfigure}{3.3in}%1.2\columnwidth}
      \includegraphics[width=\linewidth]{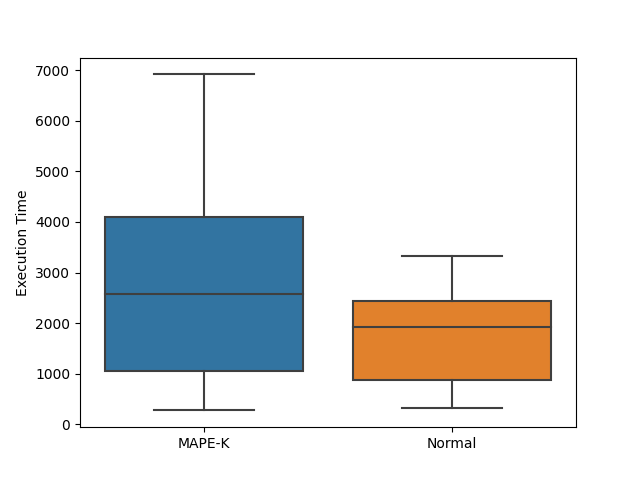}
      \caption{Total execution time between experimental treatments (seconds).}
      \label{fig:graph1}
   \end{subfigure}
\end{adjustbox}

\begin{adjustbox}{right}
   \begin{subfigure}{3.3in}%1.2\columnwidth}
      \includegraphics[width=\linewidth]{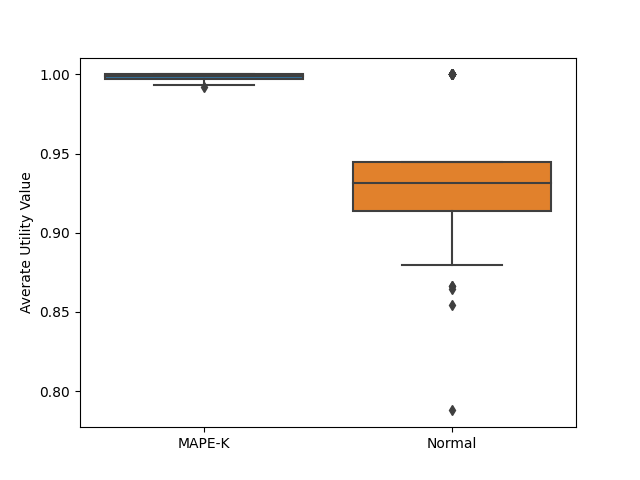}
      \caption{Average utility values for Goal (F).}
      \label{fig:graph2}
   \end{subfigure}
\end{adjustbox}
\caption{Feasibility experiments.}\label{fig:graphs}
\end{figure}

Figure~\ref{fig:graph1} presents boxplots of the overall execution time (i.e., \texttt{ticks}) for each treatment.  As can be seen by these plots, players are engaged significantly longer by the version with MAPE-K enabled ($p$ $<$ $0.05$), suggesting that the system is reconfiguring itself in support of its goals.  Figure~\ref{fig:graph2} presents the average utility values for Goal (F), where its utility equation is defined as:

\begin{equation}
   {util}_F = \begin{cases}
   1.0 \quad &\text{if} \, ps \leq w/2 \\
   0.0 \quad &\text{else if} \, ps \geq w \\
   1.0 - abs(ps - (w/2)) / (w / 2)) \quad &\text{else} \\
   \end{cases}
    \label{eqn:utilF}
\end{equation}

\noindent , where $ps$ is short for $playerSize$ and $w$ is short for the canvas width.  As can be seen in Figure~\ref{fig:graph2}, the average utility values for Goal (F) are significantly higher than those without MAPE-K ($p$ $<$ $0.05$), suggesting that adaptations can continuously improve the satisficement of Goal (F).  For the \textit{Normal} system, the utility values remain high until the player wins the game (i.e., $playerSize$ $\geq$ $width$).

%\begin{figure}[H]

Initially we had anticipated that the \texttt{FPS} would be the most significant hindrance to game execution, given the additional overhead of the p5.js library with respect to graphics processing, the large amount of entities that can spawn, and lack of interaction with a graphics processor (i.e., shaders are not compiled for this application).  The game ran at approximately $60$ \texttt{FPS} with or without MAPE-K enabled (upon visual inspection), where the main limiting factor for goal satisfaction was $playerSize$.  However, we did note that being overzealous with adaptations can in fact induce a performance decrease. 
% Unexpectedly, the \texttt{FPS} was not a significant hindrance to execution with MAPE-K enabled or disabled in this experiment.\footnote{This behavior was unexpected due to p5.js overhead.}  However, during development we noted that the additional processing required for MAPE-K \textit{can} negatively impact FPS if improperly applied.  
For example, one reconfiguration strategy was in support of Goal (E) (i.e., \textit{Maximize Alive Enemies}).  A loop was added to scale the number of active enemies as appropriate to the player's size and score but as a result the \texttt{FPS} significantly dropped and would cause violations to Goal (B) (i.e., [Maintain] 30 \texttt{FPS}), resulting in a failed execution as Goal (B) is invariant.

%\section{Experimental Results}
%\label{sec:expr}

%\section{Related Work}
%\label{sec:rw}

\section{Discussion}
\label{sec:discussion}

This paper has presented our approach for incorporating the MAPE-K feedback loop into the core gameplay loop of a single-player game.  We developed a goal model and associated utility functions for a browser-based game centered around creative coding concepts.  We additionally deployed a MAPE-K feedback loop to enable self-adaptation at run time in response to player actions and game mechanics.  Initial results indicate that applying a self-adaptive overlay to a gameplay loop can enable adaptation at run time as a result of software engineering artifacts (i.e., by elevating artifacts to be first-class citizens in the core loop).  

\noindent \textbf{Threats to Validity}:  This paper is a proof of concept to demonstrate the feasibility of run-time adaptation in the context of gaming.  As such, we have identified the following threats to validity.  For \textit{internal} validity, the derivation of all artifacts and code were performed by the authors and may be subject to missing goals or unintentional bugs.  Additionally, p5.js is not optimized for ``fast'' graphical applications and therefore suffers from bottlenecks when drawing to the canvas, where such issues are solved with more direct access (e.g., via pure JavaScript or WebGL shaders).  However, the purpose of this example is for teaching and demonstration purposes and is therefore coded in an environment that is easily approachable.   For \textit{external} validity, the configuration of the player's computer may have an impact (hardware and software combined) on the game experience, as well as the player's comfort level with gaming in general.  For \textit{construct} validity, scalability and generalizability are possible threats.  Specifically, we applied MAPE-K to a single application that is relatively simple in mechanics and limited in the number of derived goals and adaptations, however we envision that application to other types of games (both in nature and complexity) should be feasible.  
%Moreover, we focused on Goal (F) to demonstrate feasibility of our concept, however we anticipate that additional game mechanics will enable

Future paths of research for this project include human-focused studies to determine the effectiveness of self-adaptation in games (including player skill as an adaptation metric), incorporation of run-time search techniques for optimization and testing, and expansion of the \texttt{feesh} game engine for both teaching and research.

\begin{acks}
This work has been supported in part by grants from the Michigan Space Grant Consortium (\#80NSSC20M0124) and Grand Valley State University.  The views and conclusions contained herein are those of the authors and do not necessarily represent the opinions of the sponsors.
\end{acks}

\bibliographystyle{ACM-Reference-Format}
\bibliography{efredericks_master}

\end{document}